\documentclass[prl,twocolumn,showpacs,superscriptaddress,floatfix,footinbib,amsmath,amssymb]{revtex4-2}
\usepackage{graphicx}
\usepackage{amsmath}
\usepackage{cleveref}
\usepackage[usenames,dvipsnames]{color}
\usepackage[normalem]{ulem}
\usepackage{commands}
\usepackage{float}

\renewcommand{\suppl}{Supplemental Material}
\begin{document}

\title{Atomic force microscopy calibration of standing surface acoustic wave amplitudes}%

\author{Jan Hellemann}
%\email[Jan Hellemann: ]{hellemann@pdi-berlin.de}
\affiliationPDI
\author{Filipp M\"uller}
\affiliationPDI
\author{Madeleine Msall}
\affiliationPDI
\affiliationBowdoin
\author{Paulo V. Santos}
\affiliationPDI
%\author{Felix von Oppen}
%\affiliationFUtheory
\author{Stefan Ludwig}
\email[Stefan Ludwig: ]{ludwig@pdi-berlin.de}
\affiliationPDI
\date{March 2021}%
%\tableofcontents

\begin{abstract}
Atomic force microscopy is an important tool for characterizing surface acoustic waves, in particular for high frequencies, where the wavelength is too short to be resolved by laser interferometry. A caveat is, that the cantilever deflection is not equal to the amplitude of the surface acoustic wave. We show, that the energy transfer from the moving surface to the cantilever instead leads to a deflection exceeding the surface modulation. We present a method for an accurate calibration of surface acoustic wave amplitudes based on comparing force-curve measurements with the equation of motion of a driven cantilever. We demonstrate our method for a standing surface acoustic wave on a GaAs crystal confined in a focusing cavity with a resonance frequency near 3\,GHz.
\end{abstract}

\maketitle

\section{Introduction}
Today, surface acoustic waves (SAW) are used in key technologies, for instance as miniature radio-frequency filters in communication devices \cite{Ruppel2017,Mahon2017}, for sensor applications \cite{Kalinin2011,Devkota2017}, or for micro control of fluids on surfaces \cite{Wixforth2006,Ding2013}. Commonly, SAWs are generated, detected, and controlled by employing interdigital transducers, periodic arrays of metallic finger gates with a pitch of half the SAW wavelength $\lambda$. Based on the piezoelectric effect, they convert resonant electrical signals into SAWs and vice versa and also function as Bragg mirrors for SAWs. Recent research aims at integrating SAWs into quantum technology \cite{Delsing2019}, where high-quality phonon cavities that can harbor standing surface acoustic waves (SSAWs) at sub-micrometer wavelength are called for \cite{Buyukkose2013,Schuetz2015}. The cavity spectrum can be explored using the electric response of interdigital transducers \cite{Buyukkose2013}. More information, including amplitude and geometry of an SAW can be acquired by laser interferometry \cite{Santos2018}. Unfortunately, for wavelengths $\lambda<1\,\mu$m the spatial resolution of laser interferometry becomes insufficient. Nevertheless, the ability to fully characterize high-frequency SAW devices would help improving radio-frequency filters for new classical technologies \cite{Mahon2017} or, likewise, optimizing SSAW-cavities for quantum applications.

An alternative method well suited for imaging SAWs with sub-micron wavelengths is atomic force microscopy \cite{Hesjedal2009}, with one important caveat: the cantilever deflection is not a direct measurement of an SAW amplitude. The reason is the low mechanical resonance frequency of less than 1\,MHz of commonly used cantilevers, which cannot time-resolve SAWs with much larger frequencies. In this case, a cantilever interacting with an SAW can be treated as a slow oscillator driven by an external periodic force modulated with a large frequency. The transfer of kinetic energy from the quickly moving surface into the cantilever then results in an elevated average position of the cantilever tip. As a result, the measured deflection of the cantilever overestimates the SAW amplitude, independent of the detailed atomic force microscope (AFM) configuration. Here, we introduce a method to nevertheless calibrate the amplitude of an SSAW using a fixed points analysis of AFM force-curve measurements. 

\begin{figure}[t]
	\centering
    \includegraphics[width=8.6cm]{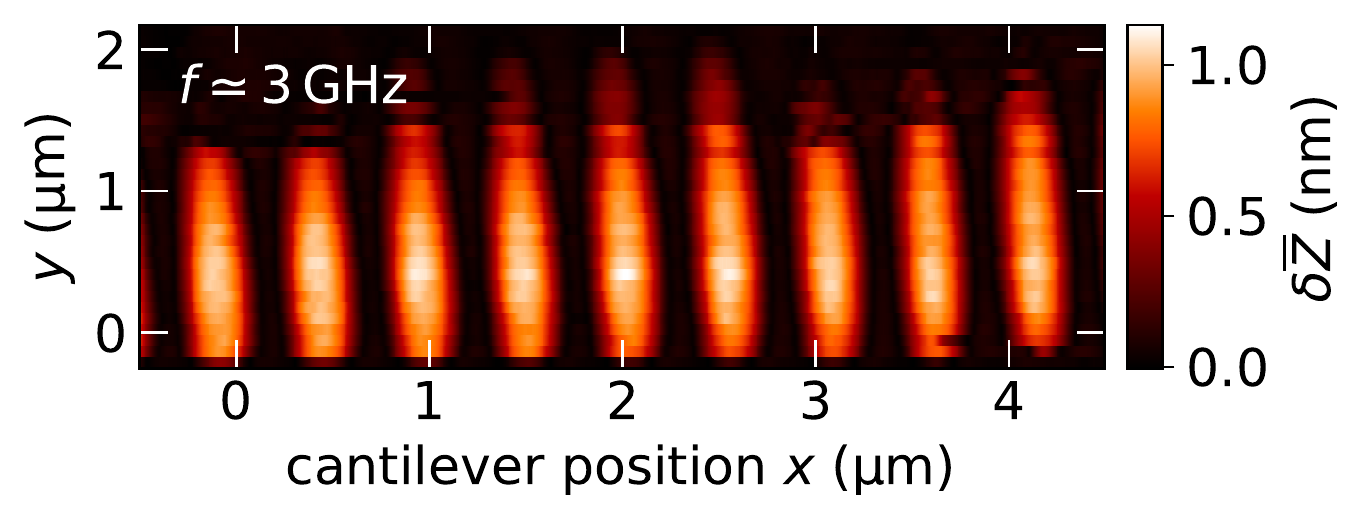}
	\caption{Two-dimensional AFM contact mode lock-in measurement of an SSAW with frequency $f\simeq3\,$GHz inside a focusing cavity on a GaAs surface using a scharp nitride cantilever with an eigenfrequency of $\omega/2\pi=65$\,kHz and a force constant of 0.32\,N/m. The excitation power was sine modulated with the lock-in frequency of $250\,$Hz between $0\le P\le9\,$dBm ($P$ is the on-chip power after subtraction of cable losses, reflections, etc.). The color scale depicts the time-averaged difference $\delta\overline Z$ of the cantilever deflection with SSAW versus without SSAW. The propagation is in the [110] crystal direction along $x$.
}
	\label{fig:AFMScan}
\end{figure}

\begin{figure}[t]
	\centering
	\vspace{-2mm}
    \includegraphics[width=8.6cm]{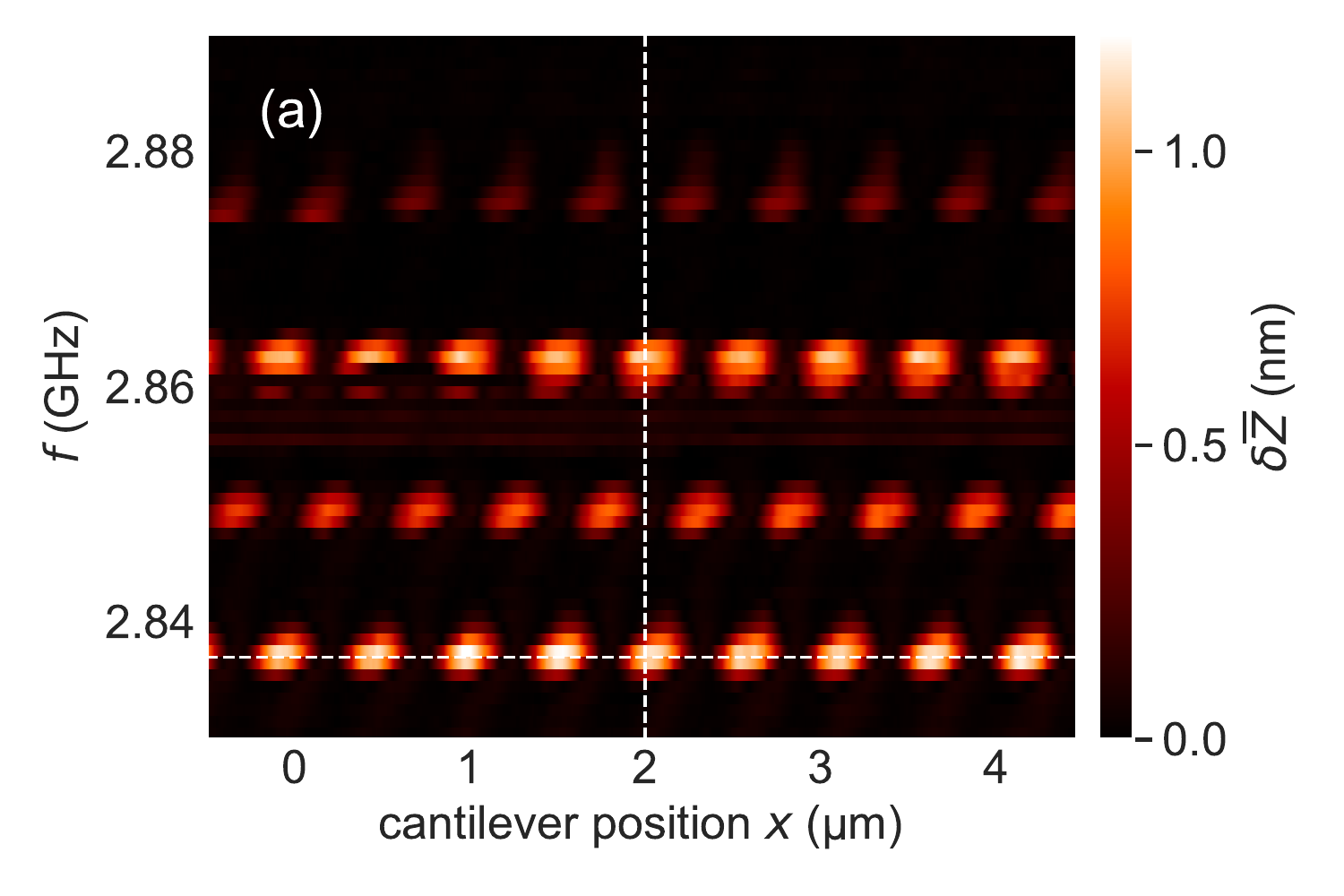} \\
    \includegraphics[width=8.6cm]{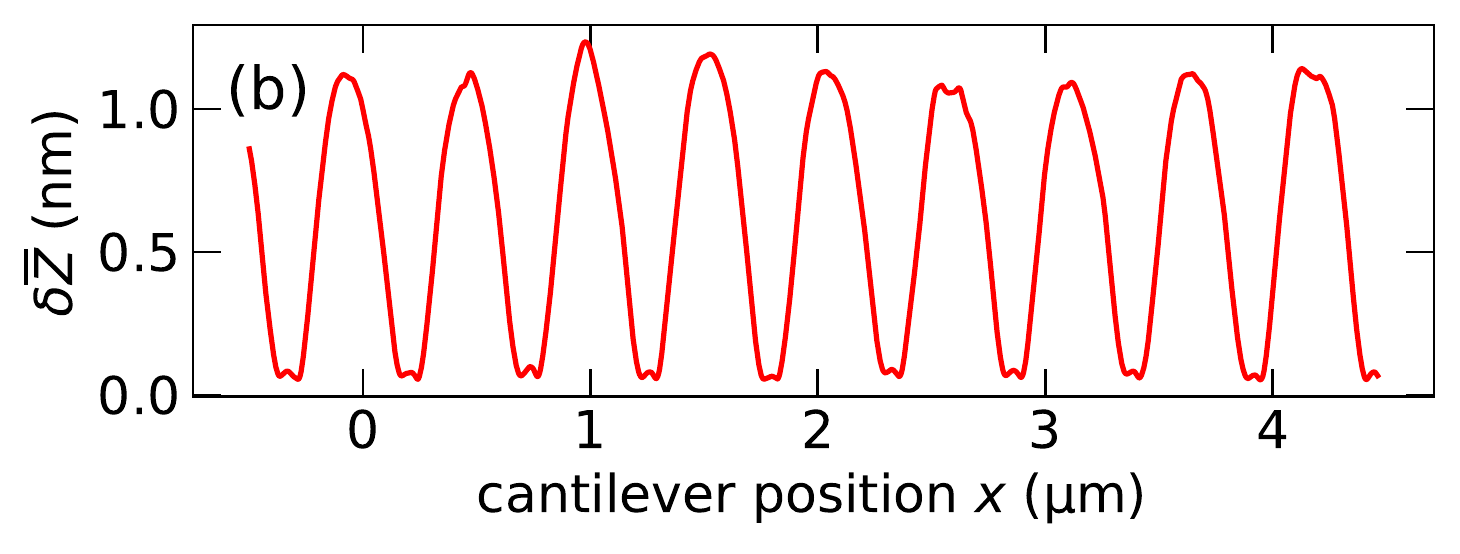}
	\caption{
	(a) AFM lock-in measurement showing $\delta\overline Z(x,f)$, details and parameters as explained in the caption of Fig.\ \ref{fig:AFMScan}. The frequency range covers four cavity modes. The vertical line is a guide for the eye, revealing phase reversal between adjacent modes.
	(b) Individual line scan $\delta\overline Z(x)$ along the horizontal line in (a) at $f=2.837\,$GHz.}
	\label{fig:requencysweep}
\end{figure}

In \fig{fig:AFMScan}{}, 
we present an extraction of an AFM measurement of a GaAs surface including an SSAW with a period of $\lambda/2=500\,$nm at a resonance frequency near $f=3\,$GHz. Clearly, the AFM measurement directly provides spatial information of an SSAW including the mode geometry and phase, which are useful for optimizing a cavity. Detecting, in addition, the frequency dependence as shown in \fig{fig:requencysweep}{a},
provides the mode spectrum of the SSAW cavity and the quality factor, here $Q\simeq1100$. Thereby, the stop bands of the cavity-defining Bragg mirrors provide the overall frequency range in which cavity modes can develop. The AFM measurement presented in \fig{fig:requencysweep}{a} includes four modes. For adjacent modes, the position of nodes and anti-nodes along the cavity is exchanged as expected at the center of the cavity [cf.\ dashed vertical line in \fig{fig:requencysweep}{a}]. The mode separation is $\delta f=12.5\,$MHz corresponding to an effective cavity length of $\lambda f\,/\,2\delta f\simeq 113\,\mu$m. In \fig{fig:requencysweep}{b}, we show a segment of an SSAW along the horizontal dashed line in \fig{fig:requencysweep}{a}. It indicates that the cantilever deflection oscillates by $\gtrsim1\,$nm between nodes and anti-nodes of this SSAW mode powered with $P=9$\,dBm.

\section{Setup and measurements}

\begin{figure}[th]
	\centering
    \includegraphics[width=8.6cm]{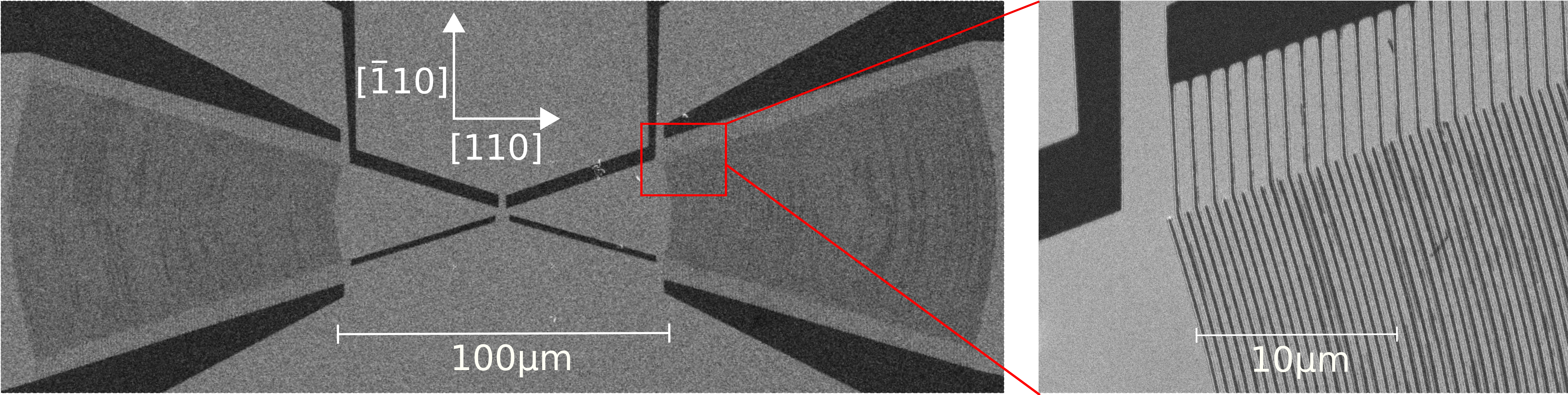}
	\caption{Scanning electron micrograph of the (001) surface (light gray) of a GaAs wafer partly covered with (10\,nm/30\,nm/10\,nm Ti/Al/Ti) metal gates (dark gray). The blow-up reveals the 250-nm-wide metal fingers at a pitch of 500\,nm forming a Bragg mirror. Two such Bragg mirrors define within the $100\,\mu$m long spacing between them a surface phonon cavity oriented along the [110] crystal direction.}
	\label{fig:schema}
\end{figure}

In \fig{fig:schema}{}, 
we present a scanning electron micrograph of the surface of our sample. It displays metal gates in dark gray on the (001) surface of a GaAs wafer shown in light gray. The blow-up reveals fingers of a Bragg mirror with a pitch of $500\,$nm for SAWs. The sample contains two matching Bragg mirrors, which together define a cavity for surface phonons in the $100\,\mu$m opening between them. While one of the Bragg mirrors is grounded, every second finger of the other Bragg mirror is connected to a radio frequency source. This way it serves as an interdigital transducer to generate an SAW. The arched fingers of the mirrors are optimized to reflect a resonant SAW such that a focused standing wave emerges inside the cavity \cite{Santos2018}.    

\begin{figure}[t]
    \centering
	\includegraphics[width=8.6cm]{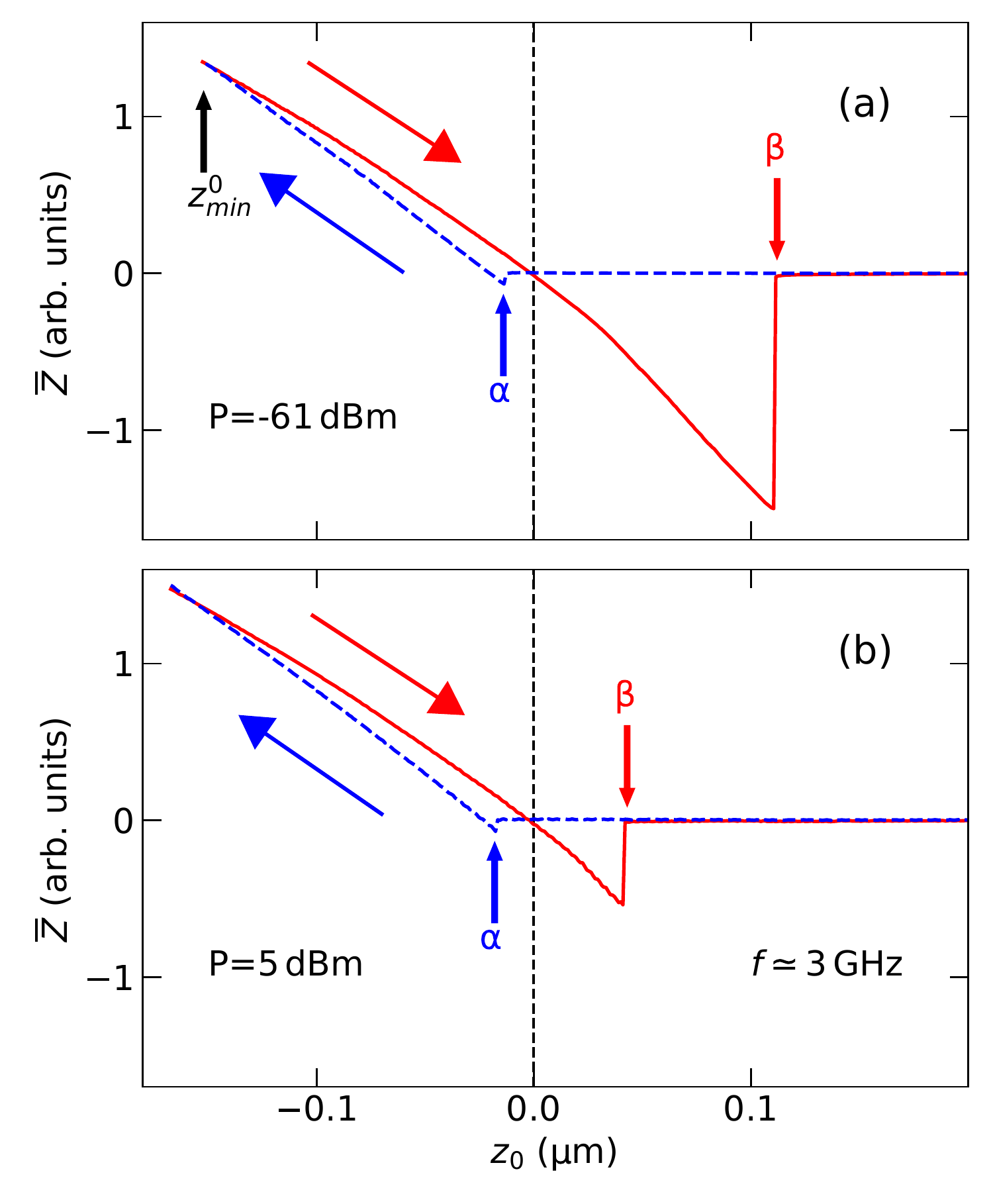}
    \caption{Force curves recorded for an anti-node of an SSAW near 3\,GHz for two different SSAW powers, $P=-61$\,dBm in (a) and 5\,dBm in (b). Plotted is the time-averaged cantilever deflection $\overline Z$ as a function of the preset cantilever position $z_0$, explained in the text. The dashed lines were measured during the approaches of the cantilever towards the surface whereas the solid lines correspond to the subsequent retractions. The position where the cantilever starts to touch the surface is indicated as $\alpha$ and the point of detachment as $\beta$. The latter strongly depends on $P$.}\label{fig:forcecurves}
\end{figure}

For recording the surface scans shown in Figs.\ \ref{fig:AFMScan}{} and \ref{fig:requencysweep} we employed the atomic force microscope \textit{Edge} by Bruker in contact mode at a constant force. However, to determine the actual SSAW amplitude, we performed force-curves: after moving the tip of the AFM cantilever to the position of an anti-node of the SSAW we measured the cantilever deflection, while vertically approaching the cantilever first towards and pushing it against the surface and then retracting it. Such a vertical motion can be accurately controlled and monitored in an AFM using a piezo-electric motor.
In \fig{fig:forcecurves}{},
we present force-curve measurements for two different SSAW powers. Plotted is the time averaged deflection $\overline Z$ of the cantilever (averaged over the resonant cantilever oscillations and many SSAW periods) as a function of its preset vertical position $z_0$, which is changed at a rate much slower than the surface modulation by the SSAW. The position $z_0$ is the distance that the tip would have from the surface if the cantilever were not deflected [cf.\ \fig{fig:tip}{}]. The dashed line was measured during the approach of the cantilever and the solid line while retracting it from the surface. As long as the cantilever moves freely above the surface its deflection remains zero during approach. A small kink of negative deflection at $z_0=\alpha$ points to an attractive force, which pulls the cantilever tip to the surface. As we continue to move the cantilever forward, the hard surface pushes its tip backwards, such that we measure an approximately linearly increasing positive deflection. The (almost) constant slope can be used for a rough calibration of the vertical AFM tip position. 

During the subsequent retraction of the cantilever (solid line) the slope of the force-curve is slightly smaller but increases with $z_0$, giving rise to a negative curvature. The hysteresis between the approach versus retraction curves in the region of positive deflection and $z_0<0$ is reproducible and commonly observed. It was interpreted in terms of a radial shirk of the cantilever \cite{Hoh1993}. Unfortunately, it conceals the accurate potential between tip and surface. For practical reasons, we define $z_0=0$ at zero average deflection of the cantilever during retraction, $\overline Z=0$. It coincides with the crossing between the retraction curve and the still horizontal approach curve. Note, that in the limit of large $z_0$ the deflection of the cantilever also vanishes.

While the cantilever is further retracted for $z_0>0$, for the time being its tip stays connected to the surface, such that $\overline Z<0$ because of the attractive force between the surface and the tip. Finally, at a height of $z_0=\beta$ the spring force of the cantilever surpasses the attraction force and the cantilever flips back to stall at $\overline Z=0$. The resulting deflection curve is characterized by a sizable triangle at negative deflection which marks a peculiar hysteresis between approach and retraction for $z_0>0$. This triangular hysteresis indicates a competition between two stable solutions.

\begin{figure}[t]
	\centering
  \includegraphics[width=8.6cm]{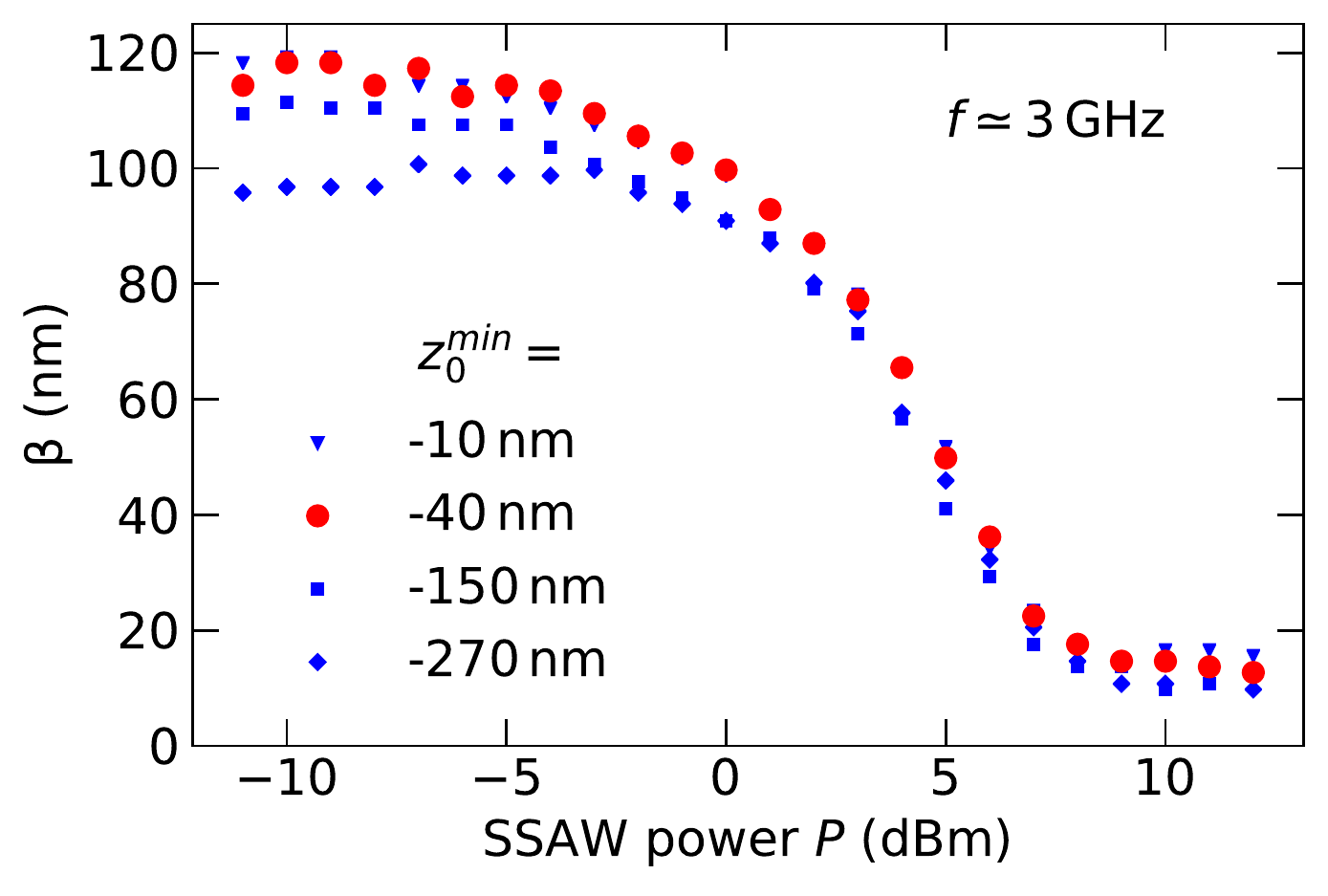}
	\caption{Position of detachment of the cantilever, $z_0=\beta$, as a function of the SSAW power $P$ for various points of return $z_0^\text{min}$, cf.\ Fig.\ \ref{fig:forcecurves}. For $z_0^\text{min}\gtrsim-150\,$nm the curves $\beta(P)$ are independent of $z_0^\text{min}$. The cantilever tip was positioned at an anti-node of an SSAW mode near 3\,GHz.
}
  \label{fig:distance_triangle_exp}
\end{figure}

In \fig{fig:forcecurves}{a} the SSAW power of $P=-61\,$dBm applied to the interdigital transducer is so small that no SSAW can be detected with the AFM, and the force-curve resembles that of an unperturbed surface. 
In \fig{fig:forcecurves}{b}, we show the same measurement but with the cantilever tip being positioned at the anti-node of a relatively strong SSAW at the frequency near $f=3\,$GHz. Strikingly the strong SSAW causes the hysteresis between approach and retraction curves, quantified as $\beta$, to be reduced. In \fig{fig:distance_triangle_exp}{}
we summarize the result of many measurements by presenting $\beta(P)$. Clearly, an increase of the SSAW power causes a smaller $\beta$. The physical reason is a reduction of the attractive force, between the cantilever tip and the surface as the cantilever tip is pushed further away by a stronger surface modulation.

For practical purposes, throughout the article we state the SSAW power $P$ estimated from the power actually applied at the generator reduced by cable losses and the electro-acoustic coupling. For the large $Q$-factor of our cavity we can neglect internal losses and estimate the SSAW power from comparing the applied electrical signal with that transmitted through the cavity. Our analysis suggests a reduction of $\simeq11\,$dB compared to the generated radio-frequency signal. Note, that the exact value of this reduction factor is irrelevant for the determination of the SSAW amplitude below.

A conversion of the SSAW power $P$ to its amplitude $R$ requires the knowledge of the function $R(P)$. While $P\propto R^2$ is evident for small powers, the proportionality factor is a-priori unknown and depends on material parameters and the exact knowledge of the power reduction factor discussed above. Our model below offers an alternative for the determination of the SSAW amplitude $R$ without requiring knowledge of either the power reduction factor or the proportionality factor. However, our model assumes the validity of $P\propto R^2$, which can be controlled in experiment and is the case for small enough amplitudes.

\section{Model and analysis}

For a quantitative analysis of the measured $\beta(P)$ and to determine the SSAW amplitudes, we introduce a model describing the dynamics of the cantilever interacting with the surface modulated by an SSAW.

\begin{figure}[t]
	\centering

	\includegraphics[width=5cm]{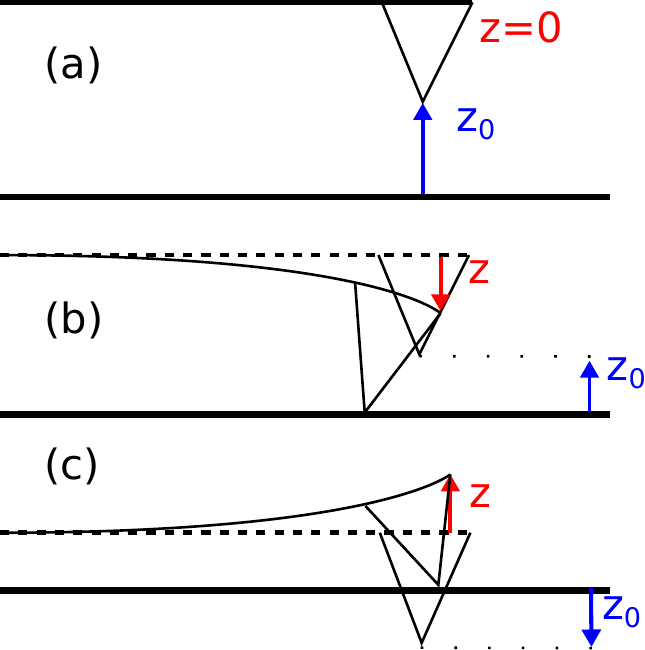}
	\caption{Sketch of the experiment illustrating various characteristic cantilever positions for $R=0$ (no SSAW). The deflection of the cantilever is $z$, the distance between surface and the tip at $z=0$ is $z_0$.  
	(a) A free cantilever $z=0$ and $z_0>0$,
	(b) a cantilever pulled towards the surface with $-z>z_0>0$,
	(c) a cantilever bent away from the surface with $-z\simeq z_0<0$.
	}
  \label{fig:tip}
\end{figure}

The equation of motion (EOM) of the cantilever can be derived from Newton's third law, $-F_\text{cl}=F_\text{i}$, where $F_\text{cl}$ is the cantilever spring force and $F_\text{i}$ the interaction force between the tip of the cantilever and the surface. It can be written as
\begin{equation}\label{eq:EOM}
 m\left(\ddot{z}+\omega^2z\right)=-V'\left[z+z_0+R\cos(\Omega t)\right]\,.
\end{equation}
Here, $z(t)$ denotes the momentary cantilever deflection and $V$ the interaction potential with $F_\text{i}=-V'=-{\partial V}/{\partial z}$. The cantilever is characterized by its force constant and its resonance frequency, we have $m\omega^2\simeq0.32\,$N/m and $\omega/2\pi\simeq65\,$kHz, respectively (manufacturer specifications). The momentary distance between the tip and the surface is $z+z_0+R\cos(\Omega t)$. The last term thereby describes the modulation of the surface by the SSAW with amplitude $R$ and frequency $f=\Omega/2\pi$. In \figs{fig:tip}{a}{c},
we sketch the cantilever set-up and indicate $z$ and $z_0$ for three relevant situations. Table III in the \suppl\ (last section in the arXiv version) provides a summary of the definitions of variables with units of length. 

\begin{figure}[t]
	\centering
  \includegraphics[width=8.6cm]{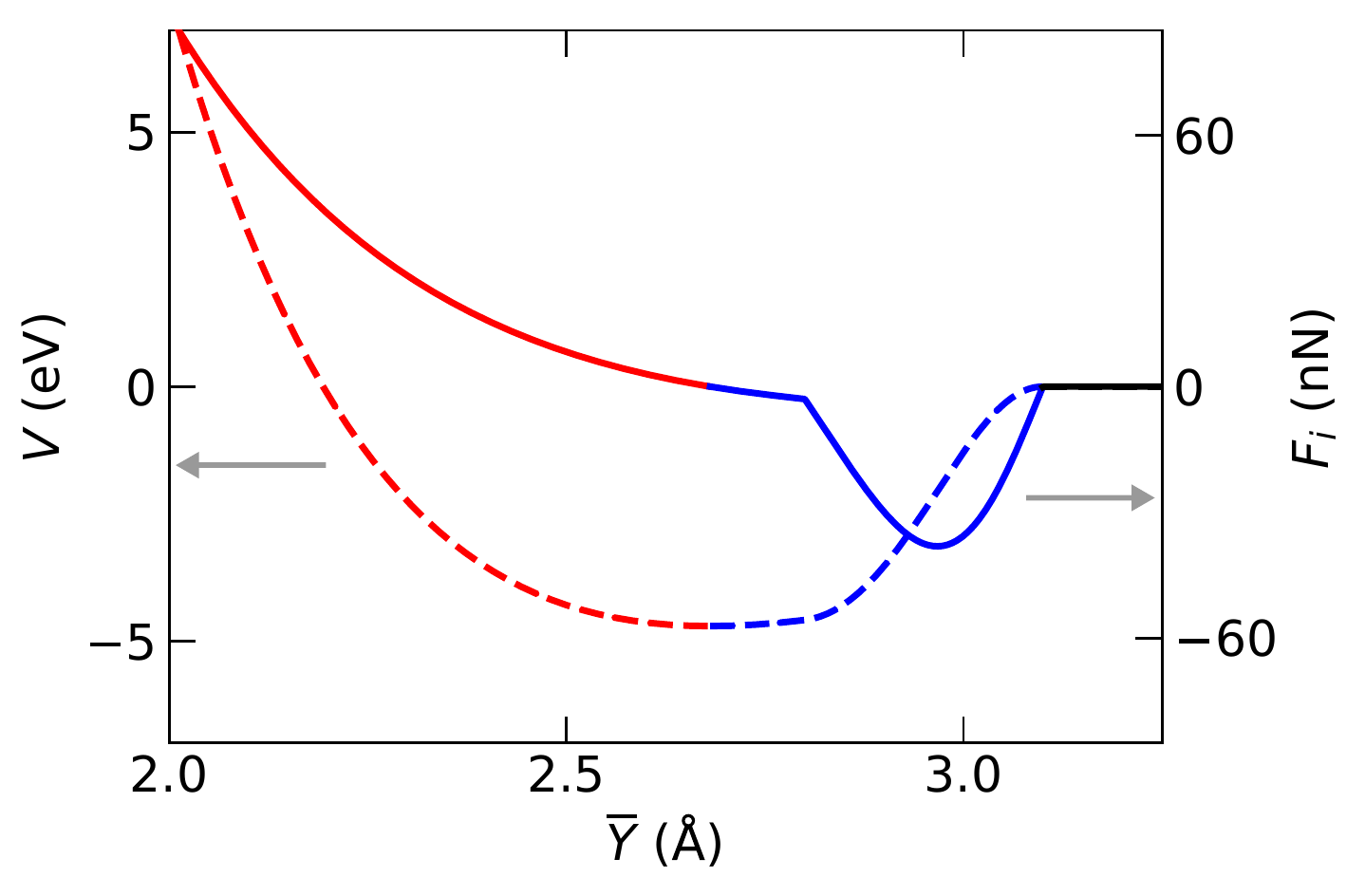}
	\caption{Optimized Tersoff potential $V$ (dashed line, left axis) and interaction force $F_\text{i}=-V'$ (solid line, right axis).
	Red color indicates regions of repulsive force $F_\text{i}>0$ and blue regions attractive force $F_\text{i}<0$. See Table II in the \suppl\ for parameters.}
  \label{fig:TersoffPotentialforce}
\end{figure}
\begin{figure*}[ht]
	\centering
	\includegraphics[width=17.0cm]{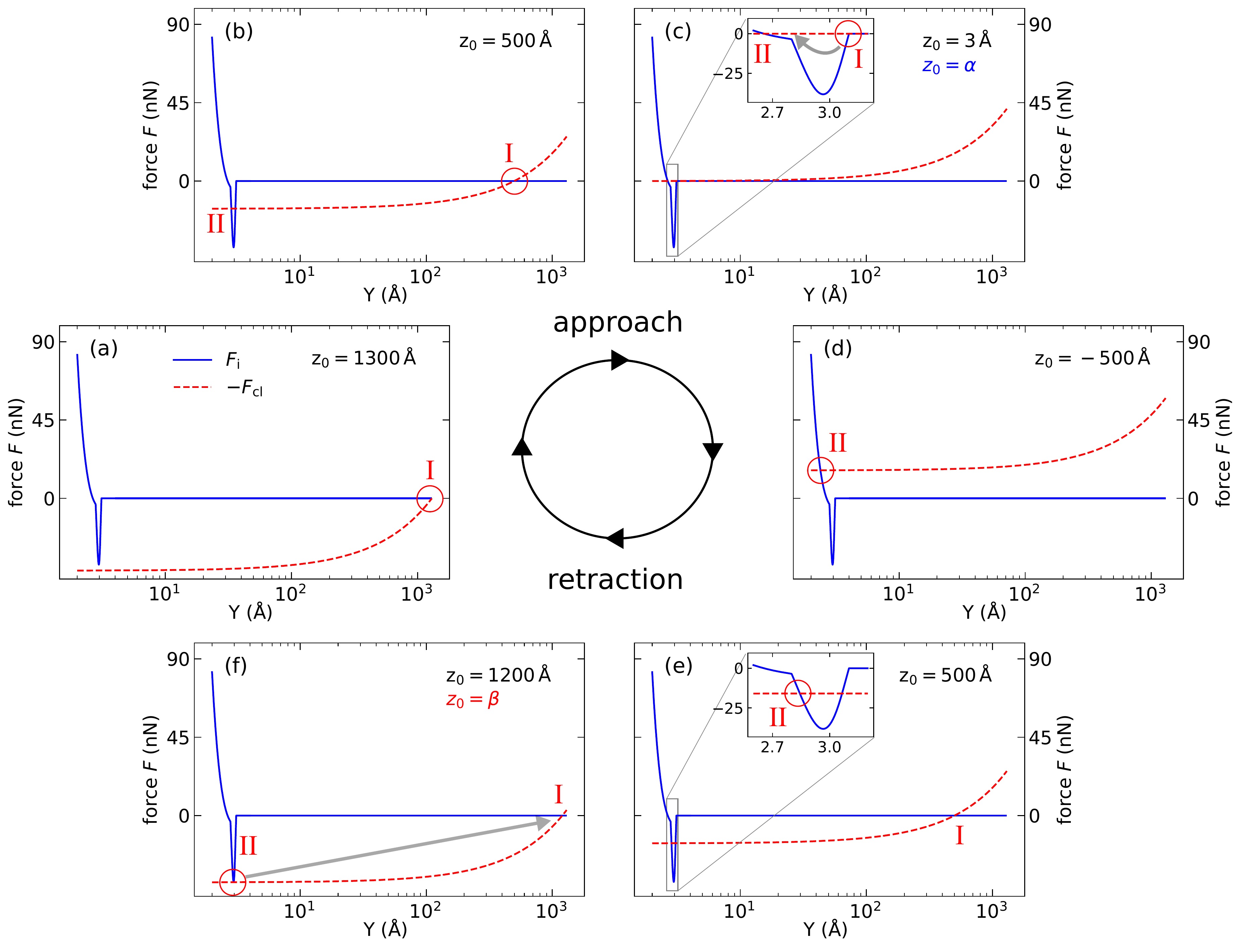}
	\caption{Solutions of the EOM for various cantilever positions during approach [$(a)\rightarrow(d)$] and retraction [$(d)\rightarrow(a)$].\\
	(a) Tip is far away from the surface, cantilever spring force $F_\text{cl}=0$, fixed point I is the only solution;
	(b) tip is approached closer to the surface, three possible fixed points, while I is occupied;
	(c) as soon as $z_0<\alpha$: transition between fixed points I $\rightarrow$ II;
	(d) tip is in contact with the surface, $z_0<\alpha$, fixed point II is the only solution;
	(e) tip remains in contact with the surface during retraction, three fixed points exist, while II is occupied;
	(f) as soon as $z_0>\beta$: transition between fixed points II $\rightarrow$ I.
}
  \label{fig:Tersoff-Cantilever}
\end{figure*}
As long as the frequency of the SSAW is orders of magnitude larger than the resonance frequency of the cantilever, in our case $\Omega/\omega>10^{4}$, we can separate the cantilever deflection into \cite{Landau1966Mech}
\begin{equation}\label{eq:timescales}
z(t)=Z(t) + \xi(t)\,,
\end{equation}
where $Z(t)$ moves slowly at a rate close to $\omega$ while $\xi(t)$ varies much faster at a rate near $\Omega$. Because the cantilever is too slow to follow the motion of the SSAW we can assume $Z\gg\xi$. We also expect $Z\gg R$ as a consequence of the transfer of kinetic energy of a hard surface to a soft cantilever. Our experimental findings confirm the latter assumption post-priori. A Taylor expansion of $V'$ around $R=0$ yields
\begin{equation}\label{eq:expansion_R}
\begin{split}
& V'\left[z+z_0+R\cos(\Omega t)\right]\simeq V'(z+z_0) +\\
& + V''(z+z_0)R\cos(\Omega t)+V'''(z+z_0)\frac{R^2}{2}\cos^2(\Omega t)\,.
\end{split}
\end{equation}
Expanding the potential derivatives on the right-hand-side of \eq{eq:expansion_R} around $\xi=0$ while neglecting all terms $\propto\xi^2$ or beyond the third derivative of $V$ we find
\begin{equation}\label{eq:expansion_xi}
\begin{split}
V'(z+z_0)   \simeq\,& V'(Y) + V''(Y)\xi\\
V''(z+z_0)  \simeq\,& V''(Y) + V'''(Y)\xi\\
V'''(z+z_0) \simeq\,& V'''(Y)\; ,
\end{split}
\end{equation}
where we introduced the slowly moving contribution of the cantilever tip distance from the surface $Y(t)=Z(t)+z_0$. Next, we split the EOM into one for the slowly varying $Z(t)$
\begin{equation}\label{eq:EOM_slow}
m(\ddot Z+\omega^2 Z)   \simeq -V'(Y)-V'''(Y)R\left[\overline{\xi \cos(\Omega t)}+\frac R4\right]
\end{equation}
and another one for the fast oscillating $\xi(t)$ 
\begin{equation}\label{eq:EOM_fast}
\begin{split}
m(\ddot\xi+\omega^2\xi) \simeq &-\bigg[V''(Y)+V'''(Y)R\cos(\Omega t)\bigg]\xi+\\
&-\left[V''(Y)+V'''(Y)\frac R2\cos(\Omega t) \right]R\cos(\Omega t)
\end{split}
\end{equation}
and insert $\xi(t)$ calculated for $\ddot\xi=-\Omega^2\xi\gg\omega^2\xi$ from \eq{eq:EOM_fast}{} into \eq{eq:EOM_slow}. Further replacing $R\overline{\xi \cos(\Omega t)}$ in \eq{eq:EOM_slow} by the leading term of its Taylor expansion around $R=0$
\begin{equation}\label{eq:taylor}
R\overline{\xi \cos(\Omega t)}=\frac{V''(\overline Y)/2}{m\Omega^2-V''(\overline Y)}R^2+{\cal O}(R^4)\,,
\end{equation}
we finally predict the average steady-state cantilever deflection (neglecting $\overline\xi\ll\overline  Z$)
\begin{equation}\label{eq:solution}
\begin{split}
\overline Z       &= -\frac{V'(\overline Y)}{m\omega^2}  +\delta\overline Z\text{ with }\\
\delta\overline Z &=- \frac{V'''(\overline Y)}{4m\omega^2}
\,\frac{m\Omega^2-3V''(\overline Y)}{m\Omega^2-V''(\overline Y)} 
\,R^2\,.
\end{split}
\end{equation}
Here, we used $\ddot Z=0$ in the steady-state and introduced the average distance between the cantilever tip and the surface $\overline Y=\overline Z+z_0$ and the average steady-state contribution of the SSAW to the deflection $\delta\overline Z\propto R^2$. The solution, \eq{eq:solution}, is resilient for $R\le0.305$\,\AA, which is the convergence radius of \eq{eq:taylor}{}. It describes the steady-state balance of forces
$-F_\text{cl}\equiv m\omega^2\overline{Z}=-V'(\overline Y) + m\omega^2\delta\overline{Z}\equiv F_\text{i}$.

The interaction potential $V$ and the amplitude $R$ of the SSAW are a-priori unknown, albeit for $R=0$ \eq{eq:solution} provides a direct relation between $V'(\overline Y)$ and the average deflection $\overline Z$. Unfortunately, the radial shirk of the cantilever while measuring the force-curve \cite{Hoh1993} leads to a strong hysteresis [cf.\ \fig{fig:forcecurves}{} for $z_0<0$]. This prevents us from determining the interaction potential with sufficient accuracy from force-curve measurements.

Nevertheless, from the force-curve for $R=0$ [cf.\ \fig{fig:forcecurves}{a}], we can estimate the attraction between the tip and the surface at $z_0=\beta=-\overline Z$ just before the cantilever detaches from the surface during retraction: $F_\text{i}(\beta)=-F_\text{cl}(\beta)=-m\omega^2\beta\simeq-40\,$nN, which points to interatomic forces holding the tip at the surface. A potential derived from interatomic forces acting between a single atom and a diamond structured semiconductor crystal was introduced by J.\ Tersoff for the case of Ge \cite{Tersoff1989}. As a practical approach, we use the Tersoff potential but slightly vary its numerical values such that it fits our data measured on a (100)-GaAs surface. In detail, we increased the attractive contribution $B$ by 1.9\,\% and the cut-off parameter $S_0$ by 0.7\,\%, while leaving all other parameters unchanged, see \suppl. The modifications are small, as expected because of the similar lattices and densities of GaAs and Ge. For completeness, in \fig{fig:TersoffPotentialforce}{}, we illustrate the optimized Tersoff potential $V$ and the corresponding interaction force $F_\text{i}=-V'$.

\begin{figure}[t]
	\centering
  \includegraphics[width=8.6cm]{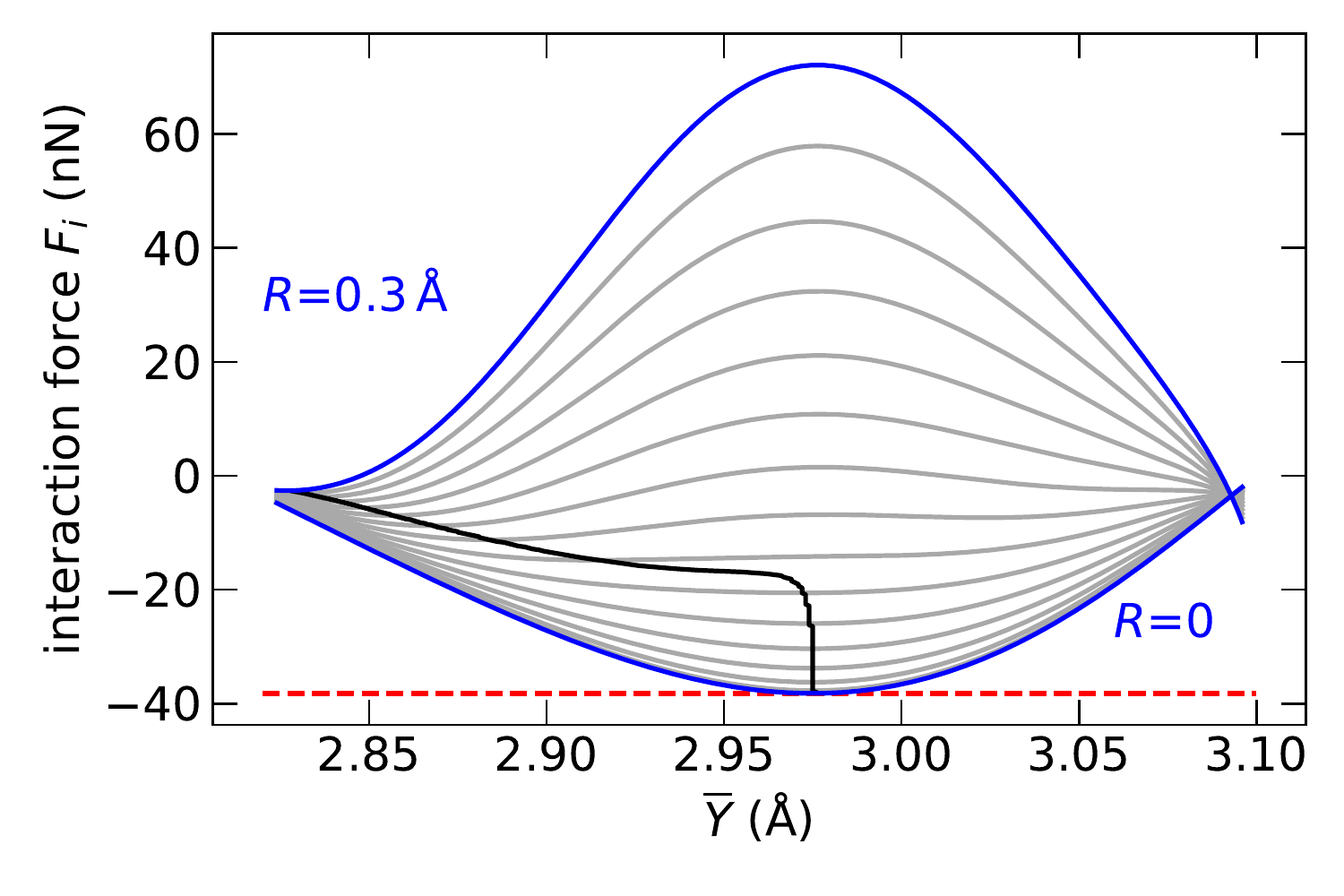}
	\caption{Interaction force $F_\text{i}$ as a function of the averaged tip-surface distance $\overline Y$ for various SSAW amplitudes $0\le R\le0.3\,$\AA\ as numerically determined by derivation of the optimized Tersoff potential shown in Fig.\ \ref{fig:TersoffPotentialforce}. The difference in $R$ between two adjacent curves is 0.02\,\AA. 
	The black line connects the points of $F_\text{i}(\overline Y)$ with slopes of 0.32\,N/m corresponding to the solutions at the cantilever detachment position $\beta$. The dashed red line is $F_\text{cl}(\overline Y)$ for $R=0$.
	}
	%The minimum of the force-curve gets more shallow the bigger the amplitude of the SAW is and therefore makes the distance smaller. The change of the minimum gets smaller for higher amplitudes. The difference between two adjacent curves is 0.02\,\AA. The black line connects the minima of the curves.}
  \label{fig:RdepGe}
\end{figure}
In \fig{fig:Tersoff-Cantilever}{} we sketch a series of snapshots for various values of $z_0$ during a force-curve measurement with $R=0$. We plot the interaction force $F_\text{i}$ (solid line) and the negative spring force of the cantilever $-F_\text{cl}=m\omega^2 Z=m\omega^2(Y-z_0)$ (dashed line) as a function of the distance $Y$ between surface and cantilever tip. The upper panels (a-d) depict the approach and the lower ones (d-a) the retraction of the cantilever. Solutions of the EOM correspond to the intersection points $F_\text{i}=-F_\text{cl}$. Depending on $z_0$, up to three solutions coexist, two of which correspond to stable fixed points. The variations between the approach versus retraction curve in \fig{fig:forcecurves}{} for $z_0>0$ are related to switches between the two stable fixed points. In \fig{fig:Tersoff-Cantilever}{} the momentarily occupied fixed point is encircled. The fixed point labeled ``I'' corresponds to a stable solution with the cantilever tip far above the surface and $Y\simeq z_0$, while fixed point ``II'' is occupied, whenever the tip is in contact with the surface and $Y$ is small (but $Z\simeq-z_0$). 

Switches between stable fixed points happen at the transition from three coexisting solutions to a single solution, corresponding to the points of exactly two solutions. During approach we observe a switch from fixed point ``I'' to ``II'' at $z_0=\alpha$ [as indicated in \fig{fig:Tersoff-Cantilever}{c}] where the attraction of the tip by the surface overcomes the cantilever spring force while $Y\simeq z_0$ and $Z\simeq0$. During retraction, the switch back from ``II'' to ``I'' happens at $z_0=\beta$ [as indicated in \fig{fig:Tersoff-Cantilever}{f}] where the cantilever spring force overcomes the attraction while $Y\ll z_0$ and $Z\simeq-z_0$.

In order to fit our model to the measured data, we use the cantilever position $z_0=\beta$ in \fig{fig:forcecurves}{} during retraction. Here, exactly two solutions exist, such that not only the forces are equal, $F_\text{i}=-F_\text{cl}$, but also their derivatives $-dF_\text{cl}/d\overline Y=dF_\text{i}/d\overline Y$, in short $-F_\text{cl}'=F_\text{i}'$, where $-F_\text{cl}'=m\omega^2=0.32\,$N/m. In \fig{fig:RdepGe}{} we graphically visualize the two conditions by plotting $F_\text{i}(\overline Y)$ while varying the SSAW amplitude within the convergence radius of our solution between $0\le R\le0.305\,$\AA. The black line, which connects the points with slope $F_\text{i}'=-F_\text{cl}'$, corresponds to the solution $\beta(R)$ with $\beta=\overline Y-\overline Z=\overline Y-F_\text{i}(\overline Y)/m\omega^2$. 

Comparing the range of $\overline Y$ in \fig{fig:RdepGe}{} with that of $\beta$ in \fig{fig:distance_triangle_exp}{}, we find that $\overline Y\ll\beta$, hence, $\overline Z\simeq-\beta$ for all our measurements. This is a direct consequence of the soft cantilever spring compared to a hard surface with the attraction force rapidly decaying with the distance $Y$ between tip and surface. It is the reason for the triangular shape of the corresponding hysteresis in \fig{fig:forcecurves}{} and leads to the observation $\delta\overline Z(z_0=\beta)\simeq\beta_\text{max}-\beta$, where $\beta_\text{max}=\beta(R=0)$. We use the measurements of $\beta(P)$ for calibrating standard AFM measurements with $z_0\ll\beta$. 

\begin{figure}[t]
	\centering
  \includegraphics[width=8.6cm]{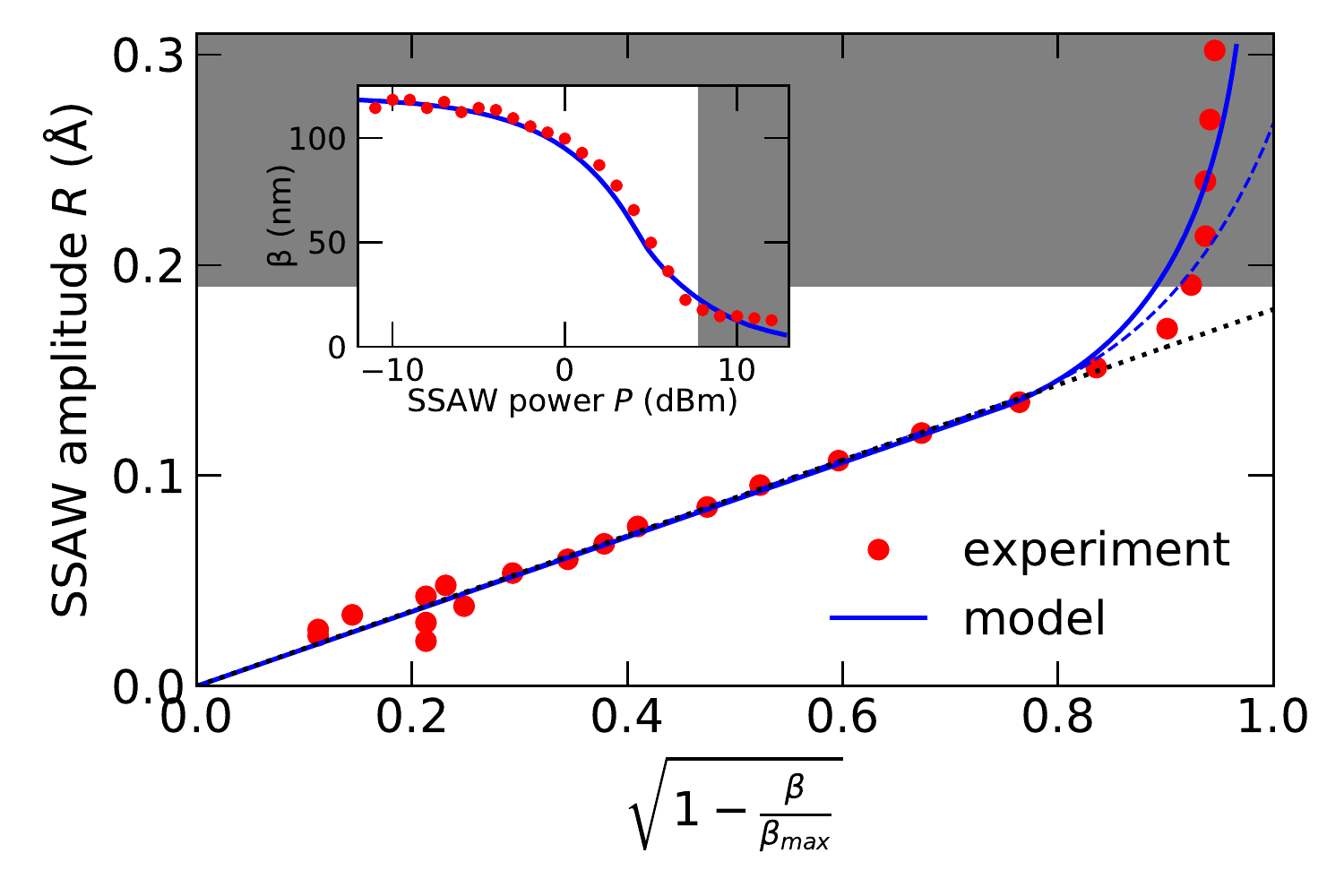}
    \caption{SSAW amplitude $R$ as a function of $\sqrt{1-\beta/\beta_\text{max}}$ determined for $z_0^\text{min}=-40$\,nm from the measured $\beta(P)$, shown in the inset and in Fig.\ \ref{fig:distance_triangle_exp}{}. The cantilever tip was positioned at an anti-node of an SSAW mode near 3\,GHz. The solid and dashed lines are model predictions, see Table II in the \suppl\ for parameters. The solid line corresponds to the potential shown in Fig.\ \ref{fig:TersoffPotentialforce}. The straight dotted line has the form $R\propto\sqrt{\beta_\text{max}-\beta}$.}
  \label{fig:calibration}
\end{figure}

Our model predicting $\delta \overline Z\propto R^2$ in \eq{eq:solution} for arbitrary $z_0$ is limited to moderate powers, in particular where $R^2\propto P$ is valid, too. To determine the proportionality factor $C=P/R^2$, we iteratively adjust both, the parameters of the Tersoff potential as well as $C$ until we find the best overall fit within the range of validity of our model.
In \fig{fig:calibration}{}
we present the resulting calibration curve $R$ as a function of $\sqrt{1-\beta/\beta_\text{max}}$ (points) determined from the measured $\beta(p)$ (points in the inset). Two model curves based on optimized Tersoff potentials are shown as solid and dashed lines, respectively. The straight dotted line represents $R\propto\sqrt{\beta_\text{max}-\beta}$, which states that $(\beta_\text{max}-\beta)\propto P$ for small powers and $\beta\ll\beta_\text{max}$. The parameters used for the model curves and the accuracy of the result are discussed in the \suppl.

The shaded region in \fig{fig:calibration}{} indicates that for $R\gtrsim0.2\,$\AA\ or $P\gtrsim8\,$dBm we begin to observe non-linear effects in AFM measurements of the SSAW showing as a reduction of the measured deflection, power broadening of the SSAW antinodes and frequency shifts. At these large powers $P\propto R^2$ is no longer valid. However, for smaller powers, the overall agreement between our measured data and the model curve is excellent. 

\section{Conclusions}

In atomic force microscopy, one directly measures the deflection of the cantilever. For an SAW with a frequency exceeding the resonance frequency of the cantilever by far, the measured deflection $\overline{Z}$ can be orders of magnitude larger than the actual amplitude $R$ of the surface modulation. The steady-state solution of the EOM describes the cantilever deflection $\delta\overline Z(R)$ in response to the surface acoustic wave. The deflection thereby depends on the derivatives of the interaction potential $V$, which however is unknown. As a workaround, we iteratively fit a suggestive model potential to the measured relation between the SSAW power $P$ and a characteristic length scale $\beta$ of the experimental force-curve, which marks a switch over between two stable solutions. In this way it is possible to reliably calibrate the actual amplitude $R$ of a standing surface acoustic wave, and ultimately as a function of the measured cantilever deflection. 

Based on force-curve measurements we have successfully calibrated  SSAW amplitudes within the limit of small powers with $R^2\propto P$. The cantilever deflection in contact mode, cf.\ \fig{fig:requencysweep}{b} for $P=9\,$dBm, is in our measurements 50 times larger than the actual SSAW amplitude.

\section{Acknowledgements}
The authors thank F.\ von Oppen and M.\ Hanke for helpful discussions, C.\ Hermann for technical support with the Bruker AFM \textit{Edge} and W.\ Anders, and A.\ Tahraoui for supporting the sample fabrication. This work was financially supported by the Deutsche Forschungsgemeinschaft through grants SA 598/15-1 and LU 819/11-1.

\section{Supplemental Material: \\Optimized Tersoff potential}

J.\ Tersoff derived empirical interatomic potentials for pure C, Si and Ge crystals and also demonstrated that only tiny modifications of these potentials are necessary to describe multicomponent systems such as SiGe \cite{Tersoff1989}. Here we start from the Tersoff potential for Ge and modify it to describe the interaction between a GaAs surface and a nitride cantilever tip. While the lattice parameters of Ge and GaAs are similar it is still surprising, how well this approach works.

The Tersoff potential is defined as \cite{Tersoff1989}
\begin{equation}\label{eq:Tersoff}
\begin{split}
V_\text{Ters}&=q f_C(Y)(A\exp(-\nu Y)-bB\exp(-\mu Y))\\
f_C(Y)&=\begin{cases} 1, &Y<S_0\\
\frac{1}{2} + \frac{1}{2}\cos[\pi\frac{Y-S_0}{S_1-S_0}], &S_0<Y<S_1\\
0, &Y>S_1
\end{cases}\\
b&=[1+\beta^n(f_C(Y)g(\theta))^n]^{\frac{-1}{2n}}\\
g(\theta)&= 1 + \frac{c^2}{d^2} - \frac{c^2}{d^2+(h-\cos(\theta))^2}\,.
\end{split}
\end{equation}
\begin{table}[thb]
\centering
\begin{tabular}{|p{17mm}|p{17mm}|p{13mm}|p{13mm}|p{15mm}|}
\hline
parameter of \phantom{abst} Tersoff potential& original value \phantom{abst} for Ge & optional modification factor & scaling factor $q$ & coefficient of determination ($R^2$-score)\\
\hline\hline
$A$ (eV)              & 1769                  & 0.981  & 4.195	& 0.893    \\
$B$ (eV)              &  419                  & 1.019  & 4.12	& 0.892    \\ 
$\nu$ (\AA$^{-1}$)    &  2.4451               & 1.0032 & 4.175	& 0.906    \\ 
$\mu$ (\AA$^{-1}$)    &  1.7047               & 0.995  & 4.065	& 0.913    \\
$\beta$               & 9.0166$\cdot$10$^{-7}$& 0.914  & 4.205	& 0.947    \\
$n$                   & 0.75627               & 1.04   & 4.165	& 0.923    \\
$c$                   & 106430                & 0.955  & 4.2	& 0.949    \\
$d$                   & 15.652                & 1.025  & 4.2	& 0.957    \\
$h$                   & -0.43884              & 0.885  & 4.205	& 0.947    \\
$S_0$ (\AA)           & 2.8                   &        &        &          \\ 
$S_1$ (\AA)           & 3.1                   &        &        &          \\
$\theta$              & $\pi/4$               &        &        &          \\
q                     & 1                     &        &        &          \\
\hline
\end{tabular}
\caption{Parameters of the Tersoff potential, original values determined for Ge and optional modification factors together with the scaling factor $q$ and the coefficient of determination. Only one of the parameters is modified at a time to adopt the potential.}
\label{table}
\end{table}
The parameters derived for Ge are listed in Table \ref{table}. To fit our data measured with a sharp nitride cantilever on a GaAs surface it is sufficient to modify a small number of these parameters. To determine the interaction between the cantilever tip and the surface, the Tersoff potentials between the tip and all individual atoms on the surface have to be summed up. We reduce this three-dimensional problem to a one-dimensional calculation by reducing the tip to a single atom and making use of the symmetry of the (001)-GaAs surface. We account for the interaction between the tip centered between four surface atoms by multiplying the Tersoff potential in \eq{eq:Tersoff} with a scaling factor $q\sim 4$. Within a self-consistent fitting procedure, we then optimize the values of $q$ such that a single modified Tersoff potential reproduces the measured interaction force of $F_\text{i}=40\,$nN for $R=0$. In the next step we optimize the factor $C=P/R^2$ and selected parameters of the Tersoff potential to achieve the best possible agreement between the measured data $\beta(P)$ and the predicted curve $R(\beta)$.

\begin{figure}[th]
	\centering
  \includegraphics[width=8.6cm]{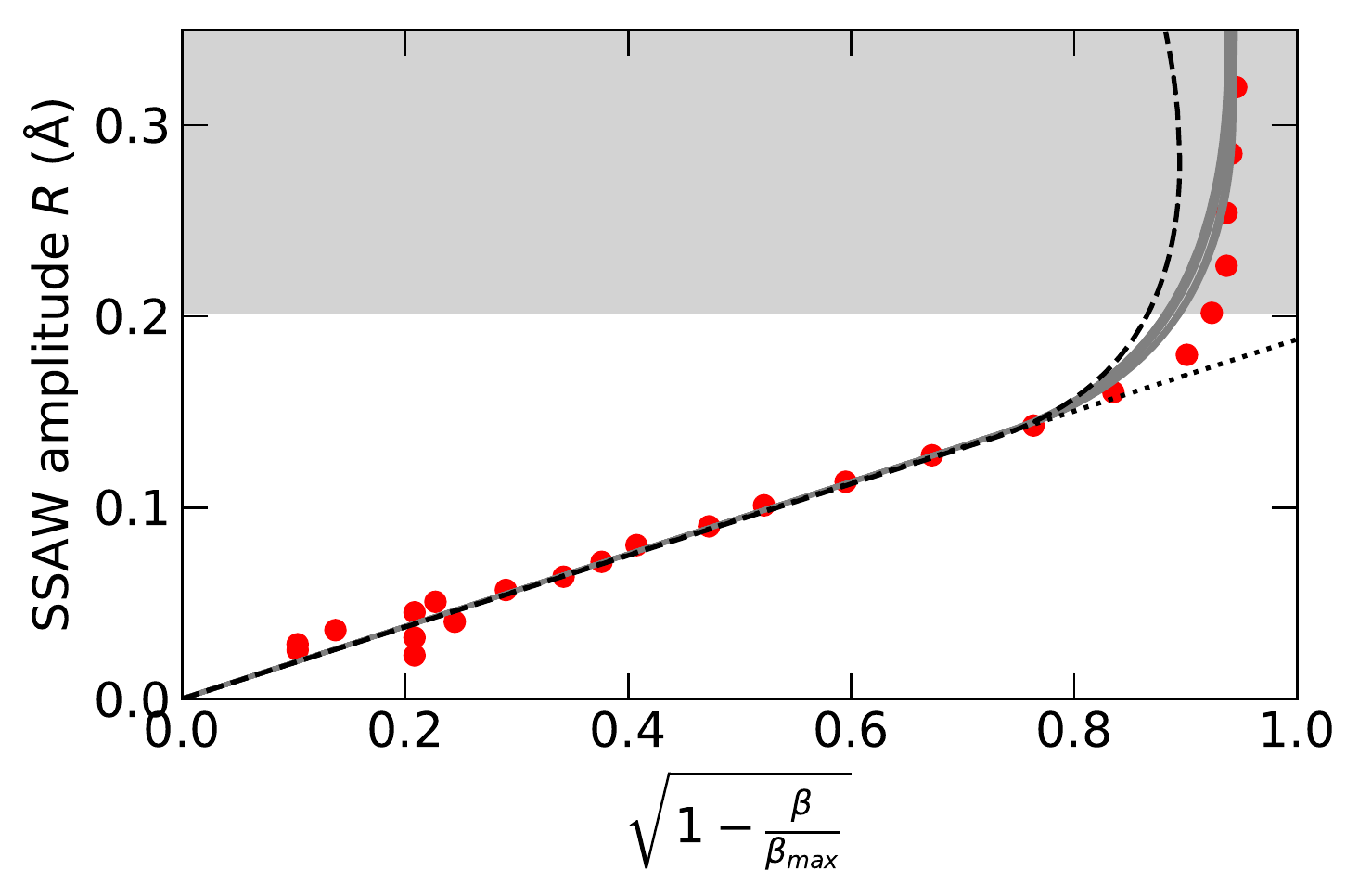}
	\caption{Calibration data similar as in Fig.\ 10 of the main article (symbols) in comparison with the model predictions (gray lines), each for one variable modified according to Table \ref{table}. The curve shown in dashed black corresponds to the unmodified Tersoff potential. The dotted curve is a straight line through the origin.}
  \label{fig:different_variables}
\end{figure}

To achieve a fair agreement, it is sufficient to slightly modify just one of the 12 parameters of the original Tersoff potential for Ge. In \fig{fig:different_variables}{}
we plot nine example model curves (gray lines) together with the experimental calibration data as in Fig.\ 10 of the main article. The dashed black line is calculated from the original Tersoff potential for Ge. In Table \ref{table} we list the optional modification factors, $q$, and the coefficient of determination ($R^2$-score). The conversion factor is $C=123$\,mW/\AA$^2$. The variations between the gray lines are relatively small and occur mainly for high powers, where our model prediction stops being accurate. For $\beta\ll\beta_\text{max}$, even the original Tersoff potential agrees well with our measurements.

\begin{table}[ht]
\centering
\begin{tabular}{|p{17mm}|p{17mm}|p{13mm}|p{13mm}|p{15mm}|}
\hline
parameter of \phantom{abst} Tersoff potential& original value \phantom{abst} for Ge & modified value used & scaling \hbox{factor $q$} used& coefficient of determination ($R^2$-score)\\
\hline\hline  
&\multicolumn{3}{c}{solid line in Fig.\ \ref{fig:calibration}, Fig.\ \ref{fig:TersoffPotentialforce}}      &                \\\hline
$B$ (eV)            & 419    &  427     & 3.9   & 0.968    \\ 
$S_0$ (\AA)         & 2.8    & 2.82  & 3.9   & 0.968    \\\hline\hline
&\multicolumn{3}{c}{dashed line in Fig.\ \ref{fig:calibration}}      &                \\\hline
$B$ (eV)            & 419    & 440  &   3.635    &   0.979        \\ 
$S_0$ (\AA)         & 2.8    & 2.82  &   3.635    &  0.979         \\\hline
\end{tabular}
\caption{Parameters of the original Tersoff potential and their modifications for the model curves presented in Fig.\ 10 of the main article. Only parameters, which we actually modified, are listed.}
\label{table-1}
\end{table}
In the main article we chose to vary $B$ in conjunction with $S_0$ as a larger value of $B$ corresponds to an increased binding energy. Qualitatively, this correction is supported by a larger melting point of GaAs at $T=1511$\,K compared to that of Ge at $T=1211.4$\,K. In Table \ref{table-1} we summarize the modifications of the Tersoff potential, which yield the model lines in Fig.\ 10 of the main article. The conversion factor is $C=138\,\text{mW}/\text{\AA}^2$, which  is roughly 10\,\% larger than the one found above for varying one parameter less. This difference corresponds just about to the accuracy of a linear fit to our data for $\beta\ll\beta_\text{max}$ (dotted line), which yields the overall accuracy of our calibration result.

Finally, in Table \ref{definitions} we list and explain the variables with units of length used in the main article.
%
%\noindent\begin{minipage}{\textwidth}
\begin{table}[ht]
\centering
\begin{tabular}{|p{30mm}|p{40mm}|}
\hline
 variable  & explanation\\\hline\hline
 $z_0$                  & distance between tip and surface for zero deflection\\
 $z(t)=Z(t)+\xi(t)$     & momentary deflection of cantilever\\
 $Z(t)$                 & slowly varying contribution of $z(t)$ (rate near $\omega$)\\
 $\xi(t)\ll Z(t)$       & quickly varying contribution of $z(t)$ (rate near $\Omega$)\\
 $Y(t)=Z(t)+z_0$        & distance between tip and surface (rate near $\omega$) \\
 $\overline Z$          & deflection of cantilever averaged over time\\
 $\overline Y=\overline Z+z_0$ & averaged distance between tip and surface\\
 $\delta\overline Z=\overline Z-\overline Z(R=0)$    & contribution of $\overline Z$ caused by SSAW \\
 $R$                    & amplitude of SSAW\\
\hline
\end{tabular}
\caption{Definitions of variables with units of length.}
\label{definitions}
\end{table}
%\end{minipage}
%

\bibliography{ludwig,zitate}

\begin{thebibliography}{14}%
\makeatletter
\providecommand \@ifxundefined [1]{%
 \@ifx{#1\undefined}
}%
\providecommand \@ifnum [1]{%
 \ifnum #1\expandafter \@firstoftwo
 \else \expandafter \@secondoftwo
 \fi
}%
\providecommand \@ifx [1]{%
 \ifx #1\expandafter \@firstoftwo
 \else \expandafter \@secondoftwo
 \fi
}%
\providecommand \natexlab [1]{#1}%
\providecommand \enquote  [1]{``#1''}%
\providecommand \bibnamefont  [1]{#1}%
\providecommand \bibfnamefont [1]{#1}%
\providecommand \citenamefont [1]{#1}%
\providecommand \href@noop [0]{\@secondoftwo}%
\providecommand \href [0]{\begingroup \@sanitize@url \@href}%
\providecommand \@href[1]{\@@startlink{#1}\@@href}%
\providecommand \@@href[1]{\endgroup#1\@@endlink}%
\providecommand \@sanitize@url [0]{\catcode `\\12\catcode `\$12\catcode
  `\&12\catcode `\#12\catcode `\^12\catcode `\_12\catcode `\%12\relax}%
\providecommand \@@startlink[1]{}%
\providecommand \@@endlink[0]{}%
\providecommand \url  [0]{\begingroup\@sanitize@url \@url }%
\providecommand \@url [1]{\endgroup\@href {#1}{\urlprefix }}%
\providecommand \urlprefix  [0]{URL }%
\providecommand \Eprint [0]{\href }%
\providecommand \doibase [0]{https://doi.org/}%
\providecommand \selectlanguage [0]{\@gobble}%
\providecommand \bibinfo  [0]{\@secondoftwo}%
\providecommand \bibfield  [0]{\@secondoftwo}%
\providecommand \translation [1]{[#1]}%
\providecommand \BibitemOpen [0]{}%
\providecommand \bibitemStop [0]{}%
\providecommand \bibitemNoStop [0]{.\EOS\space}%
\providecommand \EOS [0]{\spacefactor3000\relax}%
\providecommand \BibitemShut  [1]{\csname bibitem#1\endcsname}%
\let\auto@bib@innerbib\@empty
%</preamble>
\bibitem [{\citenamefont {Ruppel}(2017)}]{Ruppel2017}%
  \BibitemOpen
  \bibfield  {author} {\bibinfo {author} {\bibfnamefont {C.~C.~W.}\
  \bibnamefont {Ruppel}},\ }\bibfield  {title} {\bibinfo {title} {Acoustic wave
  filter technology–a review},\ }\href
  {https://doi.org/10.1109/TUFFC.2017.2690905} {\bibfield  {journal} {\bibinfo
  {journal} {IEEE Transactions on Ultrasonics, Ferroelectrics, and Frequency
  Control}\ }\textbf {\bibinfo {volume} {64}},\ \bibinfo {pages} {1390}
  (\bibinfo {year} {2017})}\BibitemShut {NoStop}%
\bibitem [{\citenamefont {Mahon}(2017)}]{Mahon2017}%
  \BibitemOpen
  \bibfield  {author} {\bibinfo {author} {\bibfnamefont {S.}~\bibnamefont
  {Mahon}},\ }\bibfield  {title} {\bibinfo {title} {The 5g effect on rf filter
  technologies},\ }\href {https://doi.org/10.1109/TSM.2017.2757879} {\bibfield
  {journal} {\bibinfo  {journal} {IEEE Transactions on Semiconductor
  Manufacturing}\ }\textbf {\bibinfo {volume} {30}},\ \bibinfo {pages} {494}
  (\bibinfo {year} {2017})}\BibitemShut {NoStop}%
\bibitem [{\citenamefont {Kalinin}(2011)}]{Kalinin2011}%
  \BibitemOpen
  \bibfield  {author} {\bibinfo {author} {\bibfnamefont {V.}~\bibnamefont
  {Kalinin}},\ }\bibfield  {title} {\bibinfo {title} {Wireless physical saw
  sensors for automotive applications},\ }in\ \href
  {https://doi.org/10.1109/ULTSYM.2011.0053} {\emph {\bibinfo {booktitle} {2011
  IEEE International Ultrasonics Symposium}}}\ (\bibinfo {year} {2011})\ pp.\
  \bibinfo {pages} {212--221}\BibitemShut {NoStop}%
\bibitem [{\citenamefont {Devkota}\ \emph {et~al.}(2017)\citenamefont
  {Devkota}, \citenamefont {Ohodnicki},\ and\ \citenamefont
  {Greve}}]{Devkota2017}%
  \BibitemOpen
  \bibfield  {author} {\bibinfo {author} {\bibfnamefont {J.}~\bibnamefont
  {Devkota}}, \bibinfo {author} {\bibfnamefont {P.~R.}\ \bibnamefont
  {Ohodnicki}},\ and\ \bibinfo {author} {\bibfnamefont {D.~W.}\ \bibnamefont
  {Greve}},\ }\bibfield  {title} {\bibinfo {title} {Saw sensors for chemical
  vapors and gases},\ }\bibfield  {journal} {\bibinfo  {journal} {Sensors}\
  }\textbf {\bibinfo {volume} {17}},\ \href {https://doi.org/10.3390/s17040801}
  {10.3390/s17040801} (\bibinfo {year} {2017})\BibitemShut {NoStop}%
\bibitem [{\citenamefont {Wixforth}(2006)}]{Wixforth2006}%
  \BibitemOpen
  \bibfield  {author} {\bibinfo {author} {\bibfnamefont {A.}~\bibnamefont
  {Wixforth}},\ }\bibfield  {title} {\bibinfo {title} {Acoustically driven
  programmable microfluidics for biological and chemical applications},\ }\href
  {https://doi.org/10.1016/j.jala.2006.08.001} {\bibfield  {journal} {\bibinfo
  {journal} {JALA: Journal of the Association for Laboratory Automation}\
  }\textbf {\bibinfo {volume} {11}},\ \bibinfo {pages} {399–405} (\bibinfo
  {year} {2006})},\ \Eprint
  {https://arxiv.org/abs/https://doi.org/10.1016/j.jala.2006.08.001}
  {https://doi.org/10.1016/j.jala.2006.08.001} \BibitemShut {NoStop}%
\bibitem [{\citenamefont {Ding}\ \emph {et~al.}(2013)\citenamefont {Ding},
  \citenamefont {Li}, \citenamefont {Lin}, \citenamefont {Stratton},
  \citenamefont {Nama}, \citenamefont {Guo}, \citenamefont {Slotcavage},
  \citenamefont {Mao}, \citenamefont {Shi}, \citenamefont {Costanzo},\ and\
  \citenamefont {Huang}}]{Ding2013}%
  \BibitemOpen
  \bibfield  {author} {\bibinfo {author} {\bibfnamefont {X.}~\bibnamefont
  {Ding}}, \bibinfo {author} {\bibfnamefont {P.}~\bibnamefont {Li}}, \bibinfo
  {author} {\bibfnamefont {S.-C.~S.}\ \bibnamefont {Lin}}, \bibinfo {author}
  {\bibfnamefont {Z.~S.}\ \bibnamefont {Stratton}}, \bibinfo {author}
  {\bibfnamefont {N.}~\bibnamefont {Nama}}, \bibinfo {author} {\bibfnamefont
  {F.}~\bibnamefont {Guo}}, \bibinfo {author} {\bibfnamefont {D.}~\bibnamefont
  {Slotcavage}}, \bibinfo {author} {\bibfnamefont {X.}~\bibnamefont {Mao}},
  \bibinfo {author} {\bibfnamefont {J.}~\bibnamefont {Shi}}, \bibinfo {author}
  {\bibfnamefont {F.}~\bibnamefont {Costanzo}},\ and\ \bibinfo {author}
  {\bibfnamefont {T.~J.}\ \bibnamefont {Huang}},\ }\bibfield  {title} {\bibinfo
  {title} {Surface acoustic wave microfluidics},\ }\href
  {https://doi.org/10.1039/C3LC50361E} {\bibfield  {journal} {\bibinfo
  {journal} {Lab Chip}\ }\textbf {\bibinfo {volume} {13}},\ \bibinfo {pages}
  {3626–3649} (\bibinfo {year} {2013})}\BibitemShut {NoStop}%
\bibitem [{\citenamefont {Delsing}\ \emph {et~al.}(2019)\citenamefont
  {Delsing}, \citenamefont {Cleland}, \citenamefont {Schuetz}, \citenamefont
  {Kn\"orzer}, \citenamefont {Giedke}, \citenamefont {Cirac}, \citenamefont
  {Srinivasan}, \citenamefont {Wu}, \citenamefont {Balram}, \citenamefont
  {B\"auerle}, \citenamefont {Meunier}, \citenamefont {Ford}, \citenamefont
  {Santos}, \citenamefont {Cerda-M{\'{e}}ndez}, \citenamefont {Wang},
  \citenamefont {Krenner}, \citenamefont {Nysten}, \citenamefont {Wei{\ss}},
  \citenamefont {Nash}, \citenamefont {Thevenard}, \citenamefont {Gourdon},
  \citenamefont {Rovillain}, \citenamefont {Marangolo}, \citenamefont
  {Duquesne}, \citenamefont {Fischerauer}, \citenamefont {Ruile}, \citenamefont
  {Reiner}, \citenamefont {Paschke}, \citenamefont {Denysenko}, \citenamefont
  {Volkmer}, \citenamefont {Wixforth}, \citenamefont {Bruus}, \citenamefont
  {Wiklund}, \citenamefont {Reboud}, \citenamefont {Cooper}, \citenamefont
  {Fu}, \citenamefont {Brugger}, \citenamefont {Rehfeldt},\ and\ \citenamefont
  {Westerhausen}}]{Delsing2019}%
  \BibitemOpen
  \bibfield  {author} {\bibinfo {author} {\bibfnamefont {P.}~\bibnamefont
  {Delsing}}, \bibinfo {author} {\bibfnamefont {A.~N.}\ \bibnamefont
  {Cleland}}, \bibinfo {author} {\bibfnamefont {M.~J.~A.}\ \bibnamefont
  {Schuetz}}, \bibinfo {author} {\bibfnamefont {J.}~\bibnamefont {Kn\"orzer}},
  \bibinfo {author} {\bibfnamefont {G.}~\bibnamefont {Giedke}}, \bibinfo
  {author} {\bibfnamefont {J.~I.}\ \bibnamefont {Cirac}}, \bibinfo {author}
  {\bibfnamefont {K.}~\bibnamefont {Srinivasan}}, \bibinfo {author}
  {\bibfnamefont {M.}~\bibnamefont {Wu}}, \bibinfo {author} {\bibfnamefont
  {K.~C.}\ \bibnamefont {Balram}}, \bibinfo {author} {\bibfnamefont
  {C.}~\bibnamefont {B\"auerle}}, \bibinfo {author} {\bibfnamefont
  {T.}~\bibnamefont {Meunier}}, \bibinfo {author} {\bibfnamefont {C.~J.~B.}\
  \bibnamefont {Ford}}, \bibinfo {author} {\bibfnamefont {P.~V.}\ \bibnamefont
  {Santos}}, \bibinfo {author} {\bibfnamefont {E.}~\bibnamefont
  {Cerda-M{\'{e}}ndez}}, \bibinfo {author} {\bibfnamefont {H.}~\bibnamefont
  {Wang}}, \bibinfo {author} {\bibfnamefont {H.~J.}\ \bibnamefont {Krenner}},
  \bibinfo {author} {\bibfnamefont {E.~D.~S.}\ \bibnamefont {Nysten}}, \bibinfo
  {author} {\bibfnamefont {M.}~\bibnamefont {Wei{\ss}}}, \bibinfo {author}
  {\bibfnamefont {G.~R.}\ \bibnamefont {Nash}}, \bibinfo {author}
  {\bibfnamefont {L.}~\bibnamefont {Thevenard}}, \bibinfo {author}
  {\bibfnamefont {C.}~\bibnamefont {Gourdon}}, \bibinfo {author} {\bibfnamefont
  {P.}~\bibnamefont {Rovillain}}, \bibinfo {author} {\bibfnamefont
  {M.}~\bibnamefont {Marangolo}}, \bibinfo {author} {\bibfnamefont {J.-Y.}\
  \bibnamefont {Duquesne}}, \bibinfo {author} {\bibfnamefont {G.}~\bibnamefont
  {Fischerauer}}, \bibinfo {author} {\bibfnamefont {W.}~\bibnamefont {Ruile}},
  \bibinfo {author} {\bibfnamefont {A.}~\bibnamefont {Reiner}}, \bibinfo
  {author} {\bibfnamefont {B.}~\bibnamefont {Paschke}}, \bibinfo {author}
  {\bibfnamefont {D.}~\bibnamefont {Denysenko}}, \bibinfo {author}
  {\bibfnamefont {D.}~\bibnamefont {Volkmer}}, \bibinfo {author} {\bibfnamefont
  {A.}~\bibnamefont {Wixforth}}, \bibinfo {author} {\bibfnamefont
  {H.}~\bibnamefont {Bruus}}, \bibinfo {author} {\bibfnamefont
  {M.}~\bibnamefont {Wiklund}}, \bibinfo {author} {\bibfnamefont
  {J.}~\bibnamefont {Reboud}}, \bibinfo {author} {\bibfnamefont {J.~M.}\
  \bibnamefont {Cooper}}, \bibinfo {author} {\bibfnamefont {Y.}~\bibnamefont
  {Fu}}, \bibinfo {author} {\bibfnamefont {M.~S.}\ \bibnamefont {Brugger}},
  \bibinfo {author} {\bibfnamefont {F.}~\bibnamefont {Rehfeldt}},\ and\
  \bibinfo {author} {\bibfnamefont {C.}~\bibnamefont {Westerhausen}},\
  }\bibfield  {title} {\bibinfo {title} {The 2019 surface acoustic waves
  roadmap},\ }\href {https://doi.org/10.1088/1361-6463/ab1b04} {\bibfield
  {journal} {\bibinfo  {journal} {Journal of Physics D: Applied Physics}\
  }\textbf {\bibinfo {volume} {52}},\ \bibinfo {pages} {353001} (\bibinfo
  {year} {2019})}\BibitemShut {NoStop}%
\bibitem [{\citenamefont {B\"uy\"ukk\"ose}\ \emph {et~al.}(2013)\citenamefont
  {B\"uy\"ukk\"ose}, \citenamefont {Vratzov}, \citenamefont {van~der Veen},
  \citenamefont {Santos},\ and\ \citenamefont {van~der Wiel}}]{Buyukkose2013}%
  \BibitemOpen
  \bibfield  {author} {\bibinfo {author} {\bibfnamefont {S.}~\bibnamefont
  {B\"uy\"ukk\"ose}}, \bibinfo {author} {\bibfnamefont {B.}~\bibnamefont
  {Vratzov}}, \bibinfo {author} {\bibfnamefont {J.}~\bibnamefont {van~der
  Veen}}, \bibinfo {author} {\bibfnamefont {P.~V.}\ \bibnamefont {Santos}},\
  and\ \bibinfo {author} {\bibfnamefont {W.~G.}\ \bibnamefont {van~der Wiel}},\
  }\bibfield  {title} {\bibinfo {title} {Ultrahigh-frequency surface acoustic
  wave generation for acoustic charge transport in silicon},\ }\href
  {https://doi.org/10.1063/1.4774388} {\bibfield  {journal} {\bibinfo
  {journal} {Applied Physics Letters}\ }\textbf {\bibinfo {volume} {102}},\
  \bibinfo {pages} {013112} (\bibinfo {year} {2013})},\ \Eprint
  {https://arxiv.org/abs/https://doi.org/10.1063/1.4774388}
  {https://doi.org/10.1063/1.4774388} \BibitemShut {NoStop}%
\bibitem [{\citenamefont {Schuetz}\ \emph {et~al.}(2015)\citenamefont
  {Schuetz}, \citenamefont {Kessler}, \citenamefont {Giedke}, \citenamefont
  {Vandersypen}, \citenamefont {Lukin},\ and\ \citenamefont
  {Cirac}}]{Schuetz2015}%
  \BibitemOpen
  \bibfield  {author} {\bibinfo {author} {\bibfnamefont {M.~J.~A.}\
  \bibnamefont {Schuetz}}, \bibinfo {author} {\bibfnamefont {E.~M.}\
  \bibnamefont {Kessler}}, \bibinfo {author} {\bibfnamefont {G.}~\bibnamefont
  {Giedke}}, \bibinfo {author} {\bibfnamefont {L.~M.~K.}\ \bibnamefont
  {Vandersypen}}, \bibinfo {author} {\bibfnamefont {M.~D.}\ \bibnamefont
  {Lukin}},\ and\ \bibinfo {author} {\bibfnamefont {J.~I.}\ \bibnamefont
  {Cirac}},\ }\bibfield  {title} {\bibinfo {title} {Universal quantum
  transducers based on surface acoustic waves},\ }\href
  {https://doi.org/10.1103/PhysRevX.5.031031} {\bibfield  {journal} {\bibinfo
  {journal} {Phys. Rev. X}\ }\textbf {\bibinfo {volume} {5}},\ \bibinfo {pages}
  {031031} (\bibinfo {year} {2015})}\BibitemShut {NoStop}%
\bibitem [{\citenamefont {Santos}\ \emph {et~al.}(2018)\citenamefont {Santos},
  \citenamefont {Msall},\ and\ \citenamefont {Ludwig}}]{Santos2018}%
  \BibitemOpen
  \bibfield  {author} {\bibinfo {author} {\bibfnamefont {P.~V.}\ \bibnamefont
  {Santos}}, \bibinfo {author} {\bibfnamefont {M.~E.}\ \bibnamefont {Msall}},\
  and\ \bibinfo {author} {\bibfnamefont {S.}~\bibnamefont {Ludwig}},\
  }\bibfield  {title} {\bibinfo {title} {Acoustic field for the control of
  electronic excitations in semiconductor nanostructures},\ }\href@noop {}
  {\bibfield  {journal} {\bibinfo  {journal} {2018 IEEE International
  Ultrasonics Symposium (IUS)}\ ,\ \bibinfo {pages} {1}} (\bibinfo {year}
  {2018})}\BibitemShut {NoStop}%
\bibitem [{\citenamefont {Hesjedal}(2009)}]{Hesjedal2009}%
  \BibitemOpen
  \bibfield  {author} {\bibinfo {author} {\bibfnamefont {T.}~\bibnamefont
  {Hesjedal}},\ }\bibfield  {title} {\bibinfo {title} {Surface acoustic
  wave-assisted scanning probe microscopy{\textemdash}a summary},\ }\href
  {https://doi.org/10.1088/0034-4885/73/1/016102} {\bibfield  {journal}
  {\bibinfo  {journal} {Reports on Progress in Physics}\ }\textbf {\bibinfo
  {volume} {73}},\ \bibinfo {pages} {016102} (\bibinfo {year}
  {2009})}\BibitemShut {NoStop}%
\bibitem [{\citenamefont {Hoh}\ and\ \citenamefont {Engel}(1993)}]{Hoh1993}%
  \BibitemOpen
  \bibfield  {author} {\bibinfo {author} {\bibfnamefont {J.~H.}\ \bibnamefont
  {Hoh}}\ and\ \bibinfo {author} {\bibfnamefont {A.}~\bibnamefont {Engel}},\
  }\bibfield  {title} {\bibinfo {title} {Friction effects on force measurements
  with an atomic force microscope},\ }\href
  {https://doi.org/10.1021/la00035a089} {\bibfield  {journal} {\bibinfo
  {journal} {Langmuir}\ }\textbf {\bibinfo {volume} {9}},\ \bibinfo {pages}
  {3310} (\bibinfo {year} {1993})}\BibitemShut {NoStop}%
\bibitem [{\citenamefont {Landau}\ and\ \citenamefont
  {Lifschitz}(1966)}]{Landau1966Mech}%
  \BibitemOpen
  \bibfield  {author} {\bibinfo {author} {\bibfnamefont {L.~D.}\ \bibnamefont
  {Landau}}\ and\ \bibinfo {author} {\bibfnamefont {E.~M.}\ \bibnamefont
  {Lifschitz}},\ }\href@noop {} {\emph {\bibinfo {title} {Lehrbuch der
  Theoretischen Physik, Band I, Mechanik}}}\ (\bibinfo  {publisher}
  {Akademie-Verlag Berlin},\ \bibinfo {year} {1966})\BibitemShut {NoStop}%
\bibitem [{\citenamefont {Tersoff}(1989)}]{Tersoff1989}%
  \BibitemOpen
  \bibfield  {author} {\bibinfo {author} {\bibfnamefont {J.}~\bibnamefont
  {Tersoff}},\ }\bibfield  {title} {\bibinfo {title} {Modeling solid-state
  chemistry: Interatomic potentials for multicomponent systems},\ }\href
  {https://doi.org/10.1103/PhysRevB.39.5566} {\bibfield  {journal} {\bibinfo
  {journal} {Phys. Rev. B}\ }\textbf {\bibinfo {volume} {39}},\ \bibinfo
  {pages} {5566} (\bibinfo {year} {1989})}\BibitemShut {NoStop}%
\end{thebibliography}%

\end{document}